\documentclass[useAMS,usenatbib,fleqn,usedcolumn]{mnras}
\usepackage[british]{babel}             
\usepackage{txfonts}                    
\usepackage[T1]{fontenc}                
\usepackage{graphicx}                   

\usepackage{bm}
\usepackage{graphicx}
\usepackage{color,ulem}

\definecolor{webgreen}{rgb}{0,.5,0}
\definecolor{webbrown}{rgb}{.6,0,0}

\hypersetup{pdfauthor={Naveen Yadav},
            pdftitle={Low frequency radio observations of SN 2011dh and the evolution of its post-shock plasma properties},
            pdfkeywords={supernovae:individual (SN 2011dh), stars: mass-loss, radiation mechanisms: non-thermal, radio continuum: general,techniques: interferometric},
            bookmarksnumbered=true
            colorlinks=true,
            breaklinks=true,
            plainpages=false,
            bookmarksnumbered=true,
            bookmarksopen=true,
            bookmarksopenlevel=1,
            urlcolor=webbrown,
            linkcolor=blue,
            citecolor=webgreen
            }

\begin{document}
\title[SN 2011dh: GMRT]{Low frequency radio observations of SN 2011dh and the evolution of its post-shock plasma properties}
\author[Naveen Yadav et al.]
{Naveen Yadav$^{1,2}$\thanks{naveen.phys@gmail.com}, Alak Ray$^{1,3}$\thanks{akr@tifr.res.in},
Sayan Chakraborti$^{3,4}$\thanks{schakraborti@fas.harvard.edu}\\
$^1$Tata Institute of Fundamental Research, Homi Bhabha Road, Mumbai 400005, India\\
$^2$Department of Physics, Indian Institute of Science, Bangalore 560012, India\\
$^3$Institute for Theory and Computation, Harvard Smithsonian Center for Astrophysics, 60 Garden Street, Cambridge, MA 02138, USA\\
$^4$Society of Fellows, Harvard University, 78 Mt. Auburn Street, Cambridge, MA 02138, USA}
\maketitle

\begin{abstract}
We present late time, low frequency observations of SN 2011dh made using the
Giant Metrewave Radio Telescope (GMRT). Our observations at $325\ \rm MHz$, $610\ \rm MHz$
and $1280\ \rm MHz$ conducted between $93-421\ \rm days$ after the explosion supplement the
millimeter and centimeter wave observations conducted between $4-15 \ \rm days$ after
explosion using the Combined Array for Research in
Millimeter-wave Astronomy (CARMA) and extensive radio observations ($ 1.0-36.5\ \rm GHz$) conducted
between $16-93\ \rm days$ after explosion using Jansky Very Large Array (JVLA). We fit a synchrotron
self absorption model (SSA) to the $610\ \rm MHz$ and $1280\ \rm MHz$ radio light curves. We use it to
determine the radius ($R_{\rm p}$) and magnetic field ($B_{\rm p}$) at $173$ \& $323$ days after the explosion.
A comparison of the peak radio luminosity $L_{\rm op}$, with the product of the peak frequency $\nu_{\rm p}$
and time to peak $t_{\rm p}$ shows that the supernova evolves between the epochs of CARMA,
JVLA and GMRT observations. It shows a general slowing down of the expansion speed of the radio
emitting region on a timescale of several hundred days during which the shock is propagating through a
circumstellar medium set up by a wind with a constant mass loss parameter, $\dot M/\varv_{\rm w}$.
We derive the mass loss parameter ($A_{\star}$) based on $610\ \rm MHz$ and $1280\ \rm MHz$ radio
light curves, which are found to be consistent with each other within error limits.
\end{abstract}
\begin{keywords}
supernovae: individual (SN 2011dh) -- stars: mass-loss -- radiation mechanisms: non-thermal  --
radio continuum: general -- techniques: interferometric.
\end{keywords}

\section{Introduction}
Identifying the nature of progenitors is a key question in the study of core-collapse and thermonuclear
supernovae. Type IIb supernova are an intermediate class of core-collapse supernova between the Type
II and Type Ib supernovae \citep{1997ARA&A..35..309F}. Their spectra show a transition from Type II
supernovae (photospheric Balmer, H$_\alpha$, lines in early time spectra, near maximum brightness)
to a spectra characteristic of Type Ib supernovae (absence of broad H$_\alpha$ emission lines in
late time spectra).  The prototypical example of type IIb supernovae is SN 1993J.
Their progenitor stars are believed to have lost most of the outer Hydrogen
envelope prior to collapse.
The mechanisms which can cause the star to lose their hydrogen envelope may differ for single
stars vs stars in binaries. The binaries can have a) mass transfer with a companion star while
b) stellar winds c) pulsations and eruptions of evolved supergiants can operate in both single and binary stars.
Most models for the prototypical IIb SN 1993J were binary based \citep{1996IAUS..165..119N,1993Natur.364..507N,
1993JApA...14...53R,1993Natur.364..509P,1994A&A...281L..89U,1994ApJ...429..300W,2014ARA&A..52..487S}.
Therefore when the supernova is young the lines are formed in the thin Hydrogen envelope and as the
supernova ejecta expands with age the inner and deeper layers are exposed revealing the Helium zone.
Type IIb supernovae therefore provide a link between the progenitors of Hydrogen rich Type II
(the Hydrogen envelope is intact before collapse) and Hydrogen poor Type Ib/c supernovae
(massive stars stripped of their Hydrogen envelope).
\citealt{2010ApJ...711L..40C} have proposed that Type IIb supernovae can be sub-divided into
compact IIb (Type cIIb) with $R_{\star}\sim 10^{11}\ \rm cm$) and
extended IIb (Type eIIb) with $R_{\star}\sim 10^{13}\ \rm cm$) based on the mass-loss history and radius of
the progenitor star and properties of the forward shock. Type eIIb have smooth radio light curves and slower
shock, whereas Type cIIb have modulations in their radio light curves and high shock velocities ($\sim 0.1c$).

SN 2011dh is a Type IIb supernova which exploded in the spiral galaxy M51 (distance taken in this work
$D \sim8.4 \pm 0.6\ \rm Mpc$ \citep{1997ApJ...479..231F}). It
was discovered in the optical by A.Riou on 2011 May 31.89 UT; \citealt{2011CBET.2736....1G}
and on 2011 Jun 01.19 UT by the Palomar Transient Factory
(PTF, \citealt{2009PASP..121.1395L,2009PASP..121.1334R}); \citealt{2011ATel.3398....1S}.
Its evolution was followed in multiple wavelengths as it was a bright nearby supernova.
\citealt{2011ATel.3413....1A} suggested its classification as a Type IIb supernova on the basis of
similarities between the spectrum of SN 2011dh reported by \citep{2011ATel.3398....1S} and a
spectrum of SN 1993J and SN 2008ax.

In this work we present the Giant Metrewave Radio Telescope (GMRT) observations of SN 2011dh. We derive the
properties of the post-shock plasma based on SSA model using combination of GMRT data and other archival data.
We put together our results with published parameters based on
Combined Array for Research in Millimeter-wave Astronomy (CARMA) and
Jansky Very Large Array (JVLA, hereafter referred as VLA) data.
In section \ref{literature}, we present a brief summary of the published work on SN 2011dh and
its progenitor star.
In section \ref{sec:data} we present the radio observations and their reduction.
The salient features of the radio emission model are discussed briefly in section \ref{sec:model}. This
is followed by the results in section \ref{sec:results}.
In section \ref{sec:discussions} we discuss the long term evolution of the parameters of this SN, including the
long term mass loss from the progenitor before it exploded as obtained from CARMA, VLA and GMRT data.
\section{Previous Optical, Radio and X-ray Observations of SN 2011dh and its Progenitor}\label{literature}
SN 2011dh was discovered in radio by \citealt{2011ATel.3411....1H} on 2011 June 04.25
UT. \citealt{2013MNRAS.436.1258H} reported extensive early time (day $4$ to day $16$ after explosion)
mm ($33.6\ \rm GHz$, $44.2\ \rm{GHz}$, $93\ \rm GHz$ and $107\ \rm GHz$)
and cm ($4.8\ \rm GHz$, $5.0\ \rm{GHz}$, $7.4\ \rm GHz$, $8.5\ \rm GHz$ and $22.5\ \rm GHz$)
CARMA and VLA observations of SN 2011dh. They concluded that in order to reconcile the
radio emission model \citep{1998ApJ...499..810C} and X-ray emission model
(inverse Compton (IC) mechanism \citep{1979ApJ...230..713B,1982ApJ...259..302C})
an equipartition value \footnote{Equipartition factor ($f_{\rm eB}$) is the ratio
of fraction of post-shock energy density in relativistic electrons and post-shock magnetic
field respectively.}, $f_{\rm eB}\approx 500-1700$ is required, which leads to a shock wave
velocity of $\sim 15000 \pm 1800\ \rm km\ s^{-1}$. \citealt{2012ApJ...752...78S} argued
that the X-ray emission can be attributed to IC mechanism provided the equipartition
assumption is relaxed ($f_{\rm eB}\approx 30$ and $\epsilon_{\rm B}\approx 0.01$) which
leads to a doubling of mass loss rate. \citealt{2012ApJ...750L..40K} observed
radio emission ($>16\ \rm days$) of SN 2011dh extensively with VLA
and fitted the observations with the synchrotron self absorption model
\citep{1998ApJ...499..810C}. It is to be noted that there is a discrepancy in the
shock velocity inferred by \citealt{2012ApJ...752...78S} ($\sim 3.0 \times 10^9 \ \rm cm/s$)
and \citealt{2013MNRAS.436.1258H} ($\sim 2.1 \times 10^9 \ \rm cm/s$) assuming equipartition.
\citealt{2012ApJ...758...81M} incorporated IC cooling
to calculate the radio light curves and obtained strong constraints on the efficiency
of the electron acceleration ($\epsilon_{\rm e}$) and magnetic field amplification
($\epsilon_{\rm B})$ in the post-shock region. They obtained $\epsilon_{\rm e}<0.01$, which is
smaller by a factor of $\sim 30$ compared to the value suggested by \citealt{2012ApJ...752...78S}.
\citealt{2012ApJ...758...81M} also suggested that a single power law electron distribution
cannot explain the radio and X-ray emission together and a pre-acceleration injection population
of electrons peaking around Lorentz factor $\gamma\sim 20-30$ in addition to a power law extending
to higher energy is required. It should be noted that the results of \citealt{2013MNRAS.436.1258H} and
\citealt{2012ApJ...752...78S} are based on early time ($<16\ \rm days$) radio emission from SN 2011dh whereas
the results of \citealt{2012ApJ...750L..40K} are based on late time ($\sim 16-93\ \rm days$) radio emission.

\citealt{2012ApJ...752...78S} inferred a high forward shock velocity
$\varv_{\rm s}\sim 0.1c$ on the basis of radio observations which indicated that the
progenitor was compact. They also used the early time ($<5\ \rm days$) optical observations
to constrain the size of progenitor star. They found the progenitor star to be consistent
with a compact progenitor at the time of explosion $R_\star \sim 10^{11}\ \rm cm$ (Type cIIb)
and is dissimilar from those of Type eIIb SNe. A similar conclusion was arrived at on  the
basis of the SN's optical light curve and measurement of its photospheric
temperature \citealt{2011ApJ...742L..18A}.
A putative progenitor was detected in the pre-explosion
Hubble Space Telescope images by \citealt{2011ATel.3399....1L}, \citealt{2011ApJ...739L..37M} and
\citealt{2011ApJ...741L..28V}. \citealt{2011ApJ...739L..37M} found that the progenitor star is consistent
with an F8 supergiant star on the basis of its spectral energy distribution (SED). On the basis of comparison with
stellar evolution tracks they suggested that it corresponds to a single star at the end of
core C-burning with an initial mass of $M_{\rm ZAMS} = 13 \pm 3\ \rm M_{\sun}$.
\citealt{2011ApJ...741L..28V} found that the star's radius ($\sim 10^{13}\ \rm  cm$) is more extended
than what has been inferred for the SN progenitor by \citealt{2012ApJ...752...78S} and
\citealt{2011ApJ...742L..18A}. They speculated that the detected star is either
an unrelated star very near the position of the actual progenitor, or, more likely, the progenitor's
companion in a mass-transfer binary system. They found the position of the detected star in a
Hertzsprung-Russell (HR) diagram to be consistent with initial mass of $M_{\rm ZAMS}\sim 17-19\ \rm M_{\sun}$.
\citealt{2011ApJ...742L...4M} estimated the mass of the progenitor based on stellar population
synthesis and found that $M_{\rm ZAMS}$ to be in close agreement with the estimate of
\citealt{2011ApJ...739L..37M}. Early optical and near IR photometry and spectroscopy of SN 2011dh
showed fast evolution. \citealt{2014A&A...562A..17E,2015A&A...580A.142E} found a small amount of hydrogen
($\sim 0.01-0.04\rm \ M_{\sun}$) in the envelope, and they made a detailed comparison with the prototypical
type IIb SN 1993J.

\citealt{2012ApJ...757...31B} computed a set of hydrodynamical models and found that
a large progenitor star with $R_\star \sim 200\ \rm R_{\sun}$ is needed to reproduce the
early light curve which was consistent with the hypothesis that the detected
yellow-supergiant in the pre-explosion HST images was the progenitor star. They
also suggested that a single star evolutionary scenario for the progenitor of SN 2011dh is
unlikely. \citealt{2013ApJ...762...74B} proposed that the progenitor belongs to a close
binary ($16\ {\rm M_{\sun}}+10\ {\rm M_{\sun}}$) system on the basis of stellar evolutionary calculations
following the evolution of both stars in the system.
\citealt{2013ApJ...772L..32V} reported the vanishing of yellow supergiant
as seen in the pre explosion images which is consistent with the analysis
by \citealt{2012ApJ...757...31B} and \citealt{2013ApJ...762...74B}.
There has been considerable debate about the nature of the residual emission from the location of the
SN. Using HST observations in the F225W and F336W bands on day 1664, \citealt{2014ApJ...793L..22F} claimed that
the residual blue point source was the hot compact companion of the progenitor YSG star that has exploded in 2011
as predicted by \citealt{2013ApJ...762...74B}. This has however been contested by \citealt{2015MNRAS.454.2580M}
who argue on the basis of UV and optical HST observations at approximately the same day that the Spectral
Energy Distribution (SED) of the late time source is inconsistent with that of a stellar source although a
partial contribution to the observed UV flux from the possibly still present companion star cannot be ruled
out. The separate claims of the two groups are subject to key assumptions made in their respective analyses.
While \citealt{2014ApJ...793L..22F} assume that the UV flux originated only from the binary companion,
\citealt{2015MNRAS.454.2580M} show this is not borne out by the expected decrease of the flux at redder wavelengths
and the observed SED is contrary to such expectations. On the other hand, the analysis of \citealt{2015MNRAS.454.2580M}
is subject to the assumption that there is no significant circumstellar interaction of the SN which is contaminating
the observed SED. Although they state that the presence of significant late time flux at optical wavelengths in
SN 2011dh suggests that the UV flux is not necessarily attributable to a binary companion, a binary companion could
still be hidden in the light of the SN itself thereby constraining the progenitor mass to a lower value than has
been estimated so far. This in turn would imply that the mass transfer which stripped the progenitor star's
hydrogen envelope was not very efficient (compared to the case for SN 1993J) and most of the mass lost from the
progenitor may have been lost from the system rather than accreted on the companion. It is possible, as claimed
by \citealt{2015MNRAS.454.2580M} that a binary companion is not a prerequisite for the progenitor of SN 2011dh,
for e.g., SN 1993J which showed a significant evidence of a hot binary companion to its
progenitor whereas in the case of SN 2008ax very late HST observations confirmed the disappearance
of the original pre-supernova star and did not show any residual from a stellar remnant at the SN position.
Our GMRT observations have bearing on these issues, which we describe below.

SN 2011dh was observed in X-rays using Swift from $\sim 3-50\ \rm days$ after the explosion
\citep{2011ATel.3420....1M,2012ApJ...752...78S,2012MNRAS.427L..70C}. It was also observed by XMM-Newton
on two epochs: 2011 June 07.20 UT \& 2011 June 11.20 UT \citep{2012MNRAS.427L..70C} and by Chandra on two epochs :
2011 June 12.30 UT \citep{2011ATel.3456....1P} \& 2011 July 03.40 UT (PI: A.M. Soderberg) at early times.
Chandra looked at M51 on 2012 Sept 09 through 2012 Oct 10 (PI: K.D. Kuntz), which provided a
long exposure of SN 2011dh corresponding to $467-498\ \rm days$ after the explosion
\citep{2014ApJ...785...95M}. \citealt{2012ApJ...752...78S} reported spectral softening with time
on the basis of spectrum extracted from June 3-7 (photon index: $0.9\pm 0.3$)
and June 7-17 (photon index: $1.8\pm 0.2$) and also noted that the X-ray luminosity is
lower by a factor of $\approx 10$ compared to the well observed Type eIIb SN 1993J and SN 2001gd.
According to their analysis, synchrotron emission at forward shock and free-free emission at reverse shock
do not explain the the origin of X-ray emission. They suggested that the X-ray emission may be due to IC
emission. \citealt{2012A&A...546A..80S} detected a hard component in the XMM-Newton
spectrum taken at $\sim 7\ \rm  days$ which disappeared by $\sim 11\ \rm days$.
They suggested that the soft component in the X-ray emission can be identified as IC emission while the harder
component has its origin in the shocked circumstellar gas. \citealt{2012MNRAS.427L..70C}
have fitted the early time ($7\ \rm day$ and $11\ \rm day$) X-ray data with two
hot diffuse gas component model originating at the forward shock and reverse shock respectively and
also show the existence of a non-negligible absorption column in addition to Galactic column
density. A similar study was done for the case of SN 1993J \citep{2002ApJ...565..419U}
in which it was found that the low-temperature component has much higher column depth
than the high-temperature component. \citealt{2014ApJ...785...95M} also derived mass-loss rate
($\sim 3\times 10^{-6}\ \rm  M_{\sun}\ yr^{-1}$ for wind velocity of $\sim 20\ \rm km s^{-1}$)
of the progenitor based on the late time ($\sim 500 \ \rm days$) Chandra X-ray observations.
Note that X-rays and radio probe different parts of the SN with the reverse shock emission measure
being sensitive to ejecta density profile, in addition to that of the circumstellar medium
while the radio probes the blastwave shock's interaction with the CSM.
The estimates of \citealt{2014ApJ...785...95M} are based on very steep ($\rho_{\rm ej}\propto r^{-n},\ n\sim 20$)
ejecta density profiles, which is similar to that inferred for SN 1993J \citep{1995ApJ...455..658S,1996ApJ...461..993F}.
\section{GMRT Observations and Data Reduction}\label{sec:data}
The data presented in this work were acquired using the Giant Metrewave Radio Telescope (GMRT)
in full intensity mode and with a $32\ \rm MHz$ bandwidth at
$325\ \rm MHz$, $610\ \rm MHz$ and $1280\ \rm MHz$ between $93-421\ \rm days$ after the explosion.
Our observations supplement the millimeter ($33.6\ \rm GHz$, $44.2\ \rm{GHz}$, $93\ \rm GHz$ and $107\ \rm GHz$)
and centimeter wave ($4.8\ \rm GHz$, $5.0\ \rm{GHz}$, $7.4\ \rm GHz$, $8.5\ \rm GHz$ and $22.5\ \rm GHz$)
observations conducted between $4-15 \ \rm days$ after explosion using the CARMA and extensive
radio observations ($ 1.0-36.5\ \rm GHz$) using the VLA conducted
between $16-93\ \rm days$ after explosion. All the datasets have been
analyzed using standard techniques in Astronomical Image Processing System (AIPS). Each dataset was
manually flagged for instances of Radio Frequency Interference (RFI) and malfunctioning
baselines and subsequently the raw visibilities were calibrated using 3C286 (which is both the
flux calibrator as well as the phase calibrator). The bandpass calibrated dataset was used for imaging
using the AIPS task IMAGR. The source flux was obtained using AIPS task
JMFIT. The reported image RMS are obtained from the region surrounding the source. The details
of the observation are presented in Table~\ref{Table:data}. The errors used in the analysis have been
calculated using
\begin{equation}
 \sigma_{\rm total}^2=\sigma_{\rm I}^2+\sigma_{\rm C}^2,
\end{equation}
to take in to account calibration errors and other unknown effects. Here $\sigma_{\rm I}$ is rms noise
in the image and $\sigma_{\rm C}$ is the $10\%$ of the determined flux taken as calibration error.
The fitting results are summarized in Table~\ref{Table:fit}.
\begin{table}
\caption{Low frequency radio observations of SN 2011dh using GMRT (Giant Metrewave Radio
Telescope). The Age is calculated assuming 2011 May 31.893 (MJD 55712.8) as the explosion
date.}
\centering
 \begin{tabular}{c c c c c}
 \hline
\rm Date of & \rm Age & \rm Frequency & \rm Flux & \rm Image RMS  \\
\rm Observation (UT)   & (Days) & \rm (GHz) & \rm (mJy) & $\sigma_{\rm I}$\rm (mJy)\\\hline
2011-09-02 & 093  & 0.607 & 0.36 & 0.08 \\
2011-10-20 & 141 & 0.607 & 1.27 & 0.09 \\
2011-10-27 & 148 & 1.276 & 4.71 & 0.04 \\
2011-10-31 & 152 & 0.323 & $<$0.60 & 0.20 \\
2012-01-15 & 228 & 0.323 & 1.17 & 0.19 \\
2012-01-16 & 229 & 0.599 & 3.12 & 0.08 \\
2012-01-16 & 229 & 1.277 & 5.90 & 0.03 \\
2012-05-09 & 343 & 1.386 & 4.86 & 0.05 \\
2012-05-16 & 350 & 0.608 & 4.53 & 0.09 \\
2012-07-20 & 415 & 0.323 & 3.61 & 0.25 \\
2012-07-21 & 416 & 0.607 & 4.24 & 0.07 \\
2012-07-26 & 421 & 1.387 & 3.60 & 0.05 \\
\hline
\end{tabular}
\label{Table:data}
\end{table}
\section{Synchrotron Self Absorption Model} \label{sec:model}
Radio emission from core collapse supernovae has long been argued to be of non-thermal origin
\citep{1982ApJ...258..790C,1982ApJ...259..302C,1998ApJ...509..861F}. It originates at the forward shock
where electrons are accelerated to relativistic energies and magnetic fields are strong. These electrons radiate by
synchrotron mechanism in the post-shock amplified magnetic field. The early time, low frequency
turnover in the spectral energy density seen in many SNe can be due to synchrotron self absorption (SSA) and/or
free-free absorption (FFA) \citep{1982ApJ...258..790C,1998ApJ...499..810C}, and their relative importance in the
context of the prototypical type IIb SN 1993J has been discussed by \citealt{1998ApJ...509..861F}. We use the
SSA model to fit our GMRT radio light curves and extract model parameters. \citealt{2012ApJ...750L..40K} also
have used the SSA model to fit the early data from VLA (and CARMA) of SN 2011dh. However, while
\citealt{2012ApJ...750L..40K} fit the broad-band spectral energy distribution at specific epochs, we use the SSA model
to fit our light curves.

We model the $610\ \rm MHz$ and the $1280\ \rm MHz$ data using the synchrotron self absorption model
\citep{1998ApJ...499..810C}. The electron index $p$ (the electron energy distribution function index
$p$ is defined as:
$N(E)\propto E^{-p}$) is taken to be $2.8$ in accordance with \citealt{2012ApJ...750L..40K}. The
blast wave expands according to a power-law: $R\propto t^m$, where the value of $m$ is taken to
be: $0.87\pm 0.07$ \citep{2012ApJ...750L..40K}. The non thermal radio flux at any given frequency
can be written as \citep{1998ApJ...499..810C}
\begin{equation}
 \frac{F(t)}{F_{\rm op}}=1.582 \left(\frac{t}{t_{\rm p}}\right)^{a}
 \left\{ 1-\exp \left(-\frac{t}{t_{\rm p}}\right)^{-(a+b)}\right\}, \label{eqn:ssa}\\
\end{equation}
where,
\begin{equation}
a = 2m+\frac{1}{2};\quad b = \frac{p+5-6m}{2},
 \end{equation}
and where $t_{\rm p}$ is the age at which the optical depth becomes unity and $F_{\rm op}$ is
the corresponding value of flux density. The values of radius ($R_{\rm p}$)
and post-shock magnetic field ($B_{\rm p}$) are determined using the following formulas from
\citealt{1998ApJ...499..810C} (for $p=2.8$ and $F_{\rm p}=1.5F_{\rm op}$)
\begin{equation}
R_{\rm p}=4.1\times10^{14}f_{\rm eB}^{-0.05}
\left(\frac{f}{0.5}\right)^{-0.05}\left(\frac{F_{\rm op}}{\rm mJy}\right)^{0.47}
\left(\frac{D}{\rm Mpc}\right)^{0.95}
\left(\frac{\nu}{5\ \rm GHz}\right)^{-1}\ \rm cm, \label{eqn:radius}
\end{equation}
\begin{equation}
B_{\rm p}=1.1 f_{\rm eB}^{-0.21}
\left(\frac{f}{0.5}\right)^{-0.21}\left(\frac{F_{\rm op}}{\rm mJy}\right)^{-0.10}
\left(\frac{D}{\rm Mpc}\right)^{-0.21}\left(\frac{\nu}{5\ \rm GHz}\right)\ \rm Gauss, \label{eqn:magn}
\end{equation}
where $f$ is the fraction of supernova volume filled with synchrotron emitting plasma, $D$ is the distance
to the supernova and $f_{\rm eB}$ is the equipartition factor which is defined as:
\begin{equation}
f_{\rm eB} = \frac{\epsilon_{\rm e}}{\epsilon_{\rm B}},
\end{equation}
where $\epsilon_{\rm e}$ and $\epsilon_{\rm B}$ are the fraction of energy density in relativistic electrons and
post-shock magnetic field respectively. The wind density profile for the case of a constant mass loss
rate $\dot M$, can be written as
\begin{equation}
 \rho_{\rm w}=\frac{\dot M}{4\pi \varv_{\rm w}}r^{-2} = Ar^{-2} \ \rm g\ cm^{-3},
\end{equation}
where $\varv_{\rm w}$ is the wind velocity. \citealt{2006ApJ...651..381C} redefine $A$ in terms of $A_\star$ as:
\begin{equation}
A_{\star}=\frac{A}{5\times 10^{11}\  \rm g\ cm^{-1}}.
\end{equation}
In the circumstellar interaction model a fraction $\epsilon_{\rm B}$,
of the thermal energy of shocked medium is converted to post-shock magnetic field energy.
The energy density in magnetic field $u_{\rm B}$,
can be related to the mass loss parameter $A_{\star}$, using Equation~8 from \citealt{2006ApJ...651..381C} as
\begin{equation}
 u_{\rm B} = \frac{B^2}{8\pi}=0.052\left(\frac{\epsilon_{\rm B}}{0.1}\right)
 A_{\star}\left(\frac{t}{10\ \rm d}\right)^{-2} \ \rm erg\ cm^{-3}.
\end{equation}
The quantity $A_{\star}$ can thus be related to the observed quantities by rewriting the above equation as
\begin{equation}
 A_{\star}=0.76\left(\frac{\epsilon_{\rm B}}{0.1}\right)^{-1}\left(\frac{t}{10
\ \rm d}\right)^2B^2.
\end{equation}

\section{Results}\label{sec:results}
\begin{figure}
  \begin{center}
   \includegraphics[scale=1.0]{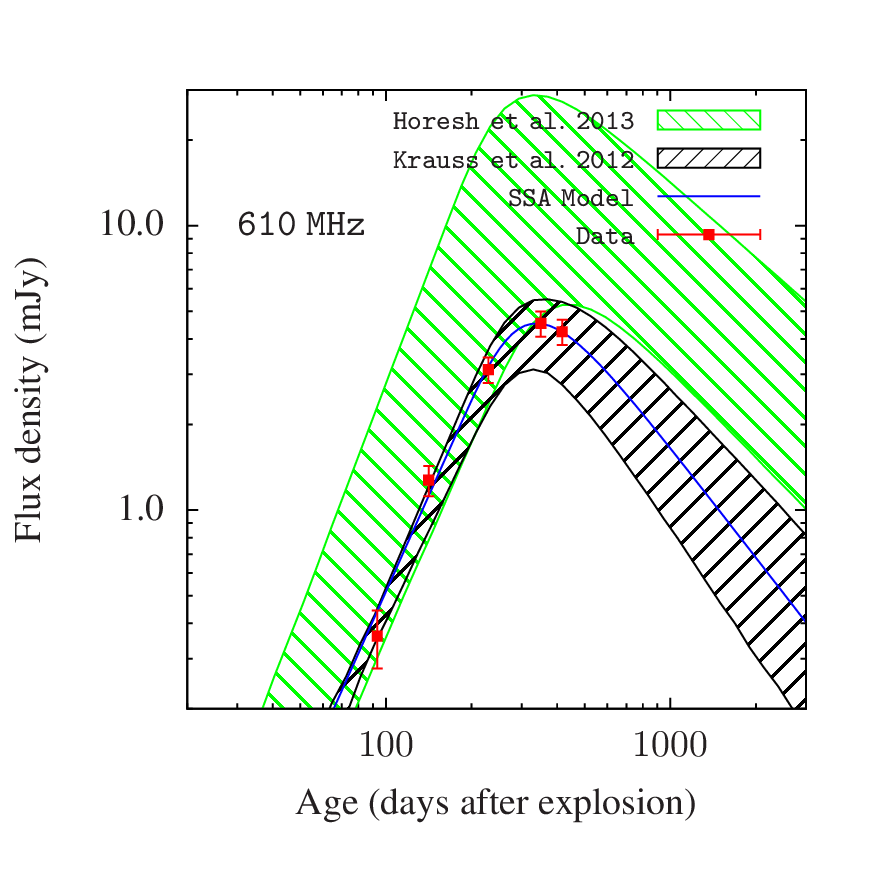}
   \includegraphics[scale=1.0]{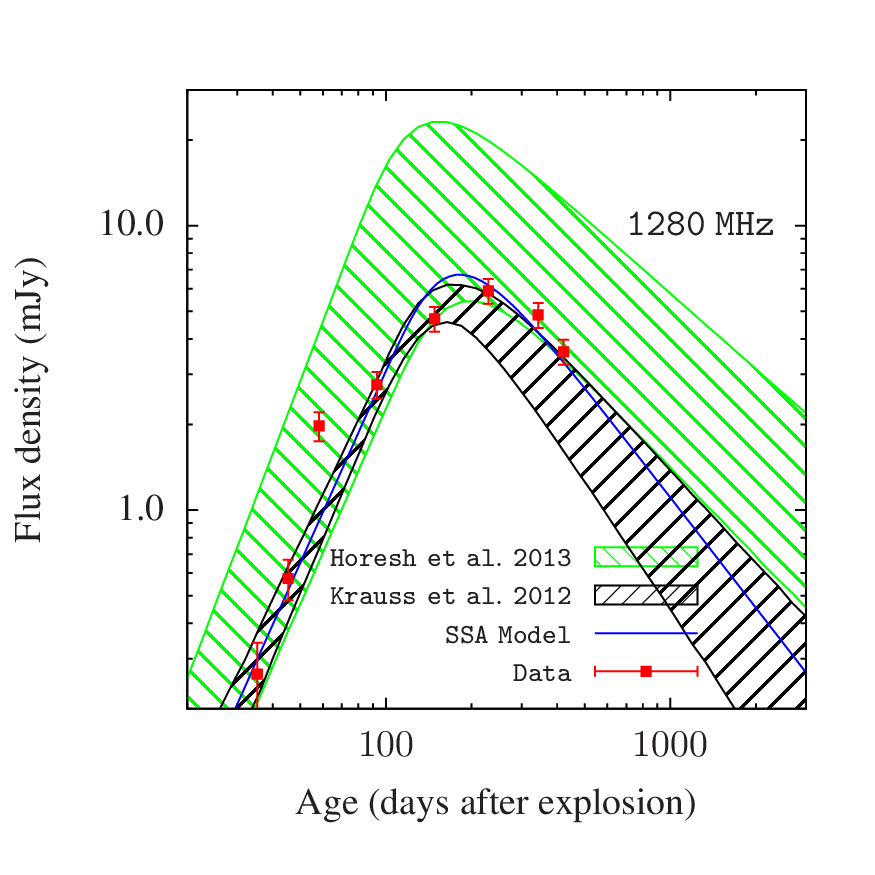}
   \caption{The $610\ \rm MHz$ data (upper panel) and $1280\ \rm MHz$ data (lower panel) of
SN 2011dh fitted with the synchrotron self absorption model (see section \ref{sec:model}) \citep{1998ApJ...499..810C}.
The curve at $1280\ \rm MHz$ consists of GMRT and VLA data \citep{2012ApJ...750L..40K}
(Here, $1.4\ \rm GHz$ flux is scaled to $1.28\ \rm GHz$ using optically thick scaling: $F\propto \nu^{5/2}$).
The best fit parameters for the fits are reported in the Table~\ref{Table:fit}. The green and black shaded regions
are predictions of the $610\ \rm MHz$ \& $1280\ \rm MHz$ fluxes made using the model and parameters reported in
\citealt{2013MNRAS.436.1258H} ($\alpha = -1.15,\ \beta =-0.96,\ K1=453.43,\ K5=1.9772\times 10^{5}\ \&\ \delta''=-3.42$, with
$1\sigma$ errors of $6,\ 8,\ 8,\ 12$ and $3\%$) and \citealt{2012ApJ...750L..40K} (at $92.9\ \rm day$
, $S_{{\nu}_{\tau}}=6.44\pm0.21\ \rm mJy$ and $\nu_{\tau}=2.235\pm0.076\ \rm GHz$) respectively.}
\label{fig:fits}
  \end{center}
\end{figure}

\begin{figure}
  \begin{center}
   \includegraphics[scale=1.0]{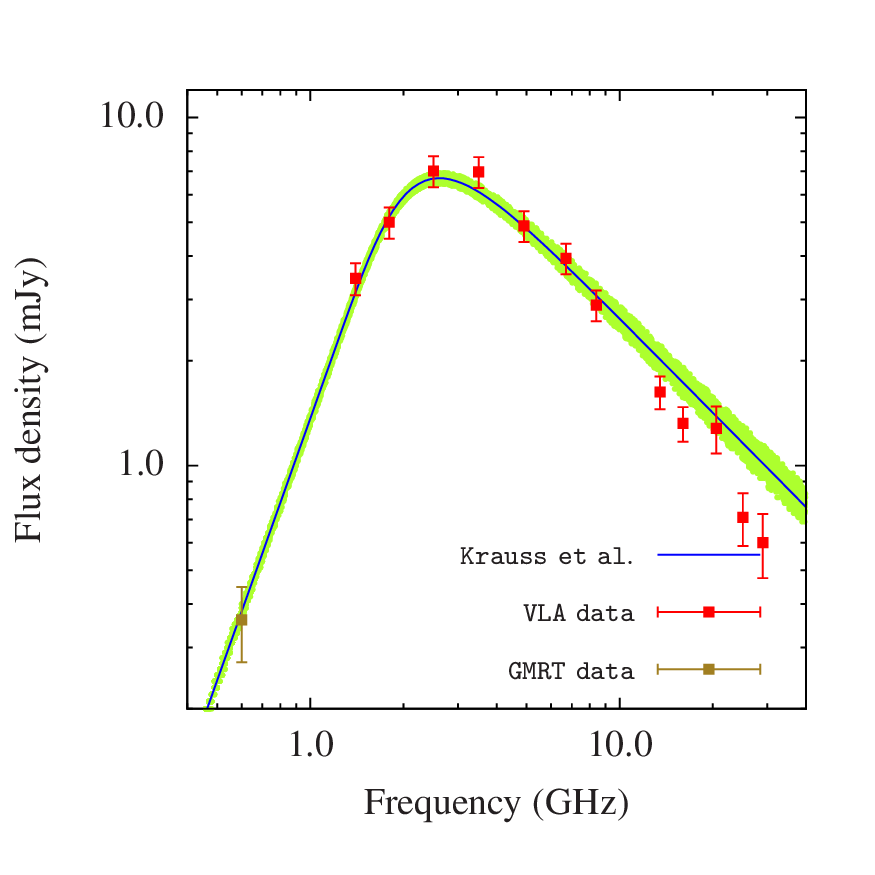}
   \caption{The $610\ \rm MHz$ GMRT data point at $93.2\ \rm day$ plotted along side the spectrum on $92.9\ \rm day$
   reported by \citealt{2012ApJ...750L..40K}. The solid curve is the best fit to the spectrum as reported by
   \citealt{2012ApJ...750L..40K} and it shows that $610\ \rm MHz$ flux density determined by GMRT is consistent with
   the SSA model fitted only to the VLA data on $92.9\ \rm day$. The green shaded region corresponds to the
   uncertainties in the fit parameters.}\label{fig:spec_92}
  \end{center}
\end{figure}

We fit the low frequency data with the synchrotron self absorption model (see Equation~\ref{eqn:ssa})
which has been presented in the previous section (fixing the value of $m$ and $p$ from \citealt{2012ApJ...750L..40K}).
The fits to the low frequency radio data are shown in Figure~\ref{fig:fits}.
The $610\ \rm MHz$ data point on $93.2\ \rm day$ falls on the spectrum reported
by \citealt{2012ApJ...750L..40K} on $92.9\ \rm day$ as shown in Figure~\ref{fig:spec_92}.
The fit at $1280\ \rm MHz$ shown in Figure~\ref{fig:fits} (bottom panel) comprises of all the available
L-Band data from GMRT and VLA \citep{2012ApJ...750L..40K}. The $1.40\ \rm GHz$ VLA data has been scaled
to $1.28\ \rm GHz$ (using optically thick synchrotron emission scaling: $F_{1.28}/F_{1.40}=(1.28/1.40)^{5/2}$)
for the purpose of fitting. We derive the values of $R_{\rm p}$ and $B_{\rm p}$ using Equation~\ref{eqn:radius}
\& \ref{eqn:magn} and assuming equipartition ($f_{\rm eB}=1$).

\begin{figure}
  \begin{center}
   \includegraphics[scale=1.0]{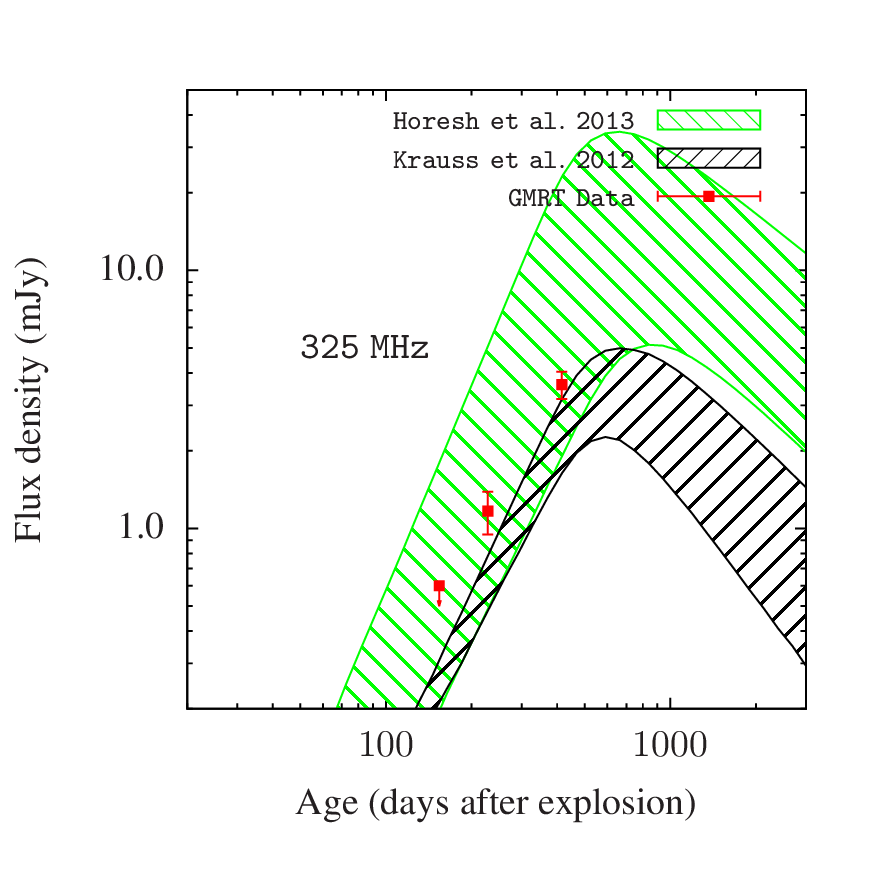}
   \caption{The $325\ \rm MHz$ data of SN 2011dh. The green and black shaded regions
are predictions of the $310\ \rm MHz$ fluxes made using the model and parameters reported
in \citealt{2013MNRAS.436.1258H} and \citealt{2012ApJ...750L..40K} respectively.}\label{fig:fits325}
  \end{center}
\end{figure}

We can use the radio emission models derived by \citep{2012ApJ...750L..40K} and \citealt{2013MNRAS.436.1258H} to
predict the radio flux densities for GMRT observations. We have plotted the predicted flux densities
based on model from \citealt{2012ApJ...750L..40K} and \citealt{2013MNRAS.436.1258H} along with the observed flux
densities in Figure~\ref{fig:fits} \& \ref{fig:fits325}. \citealt{2012ApJ...750L..40K} have used the
SSA model to fit the broadband radio spectra on various days. The
model is written as (Equation~1 in \citealt{2012ApJ...750L..40K})
\begin{equation}
 S(\nu) = 1.582S_{\nu_{\tau}}\left(\frac{\nu}{\nu_{\tau}}\right)^{5/2}
 \left\{1-\exp\left[-\left(\frac{\nu}{\nu_{\tau}}\right)^{-(p+4)/2}\right]\right\},
\end{equation}
where $S_{\nu_{\tau}}$ is the flux density at frequency $\nu_{\tau}$, at which the optical depth
is unity. The \citealt{2012ApJ...750L..40K} predictions shown in Figure~\ref{fig:fits} \& \ref{fig:fits325}
are calculated using their fit parameters for broadband radio spectra at $92.9\ \rm day$
($S_{{\nu}_{\tau}}=6.44\pm0.21\ \rm mJy$ and $\nu_{\tau}=2.235\pm0.076\ \rm GHz$). We have taken
the parameters corresponding to $92.9\ \rm day$ as it is the closest to GMRT observation epoch
on $93 \rm \ day$ and onwards. To calculate the flux densities based on \citealt{2013MNRAS.436.1258H}
we have used the following equations from \citealt{2013MNRAS.436.1258H}:
\begin{equation}
 S=K1\left(\frac{\nu}{5\ \rm GHz}\right)^{\alpha}\left(\frac{t-t_0}{1\ \rm day}\right)^{\beta}
 \left(\frac{1-e^{-\tau_{\rm ssa}}}{\tau_{\rm ssa}}\right),\label{eqn:Horesh1}
\end{equation}
\begin{equation}
 \tau_{\rm ssa}=K5\left(\frac{\nu}{5\ \rm GHz}\right)^{\alpha-2.5}\left(\frac{t-t_0}{1\ \rm day}\right)^{\delta ''},\label{eqn:Horesh2}
\end{equation}
where $K1$ and $K5$ are proportionality constants and $\delta ''$ describes the time dependence of the optical
depth (the parameters have the values $\alpha = -1.15,\ \beta =-0.96,\ K1=453.43,\ K5=1.9772\times 10^{5}\ \&\ \delta''=-3.42$,
with $1\sigma$ errors of $6,\ 8,\ 8,\ 12$ and $3\%$ , respectively.)).
We find that the predicted flux densities from \citealt{2013MNRAS.436.1258H} deviate more from the
observed late time data compared to the predicted flux densities based on the model
from \citealt{2012ApJ...750L..40K}. This may be due to the fact that the two models were based on different
time ranges of observations and the parameters that determine the radio emission
have changed with time. This is in contrast to many cases where the radio emission from a supernova can be
consistently fitted with a model valid across temporal and frequency coverage of the observations.
We note that \citealt{1998ApJ...509..861F} advocated an evolution in the shock speed for the case of SN 1993J, with
a break after $\sim 100$ days. For the case of $325\ \rm MHz$ observations, we have detection at two epochs
and an upper limit on the flux density. Therefore we have not attempted to fit the radio light curve, and we
show the observed data in Figure~\ref{fig:fits325} along with the predicted values.
We have obtained radius and magnetic field at two epochs using Equation~\ref{eqn:radius} \& \ref{eqn:magn}.
The errors in these quantities have been estimated using
\begin{equation}
 \frac{\delta R_{\rm p}}{R_{\rm p}} = \left(\left(0.47\frac{\delta F_{\rm op}}{F_{\rm op}}\right)^2+
 \left(0.95\frac{\delta D}{D}\right)^2+
 \left(\frac{\delta \nu_{\rm p}}{\nu_{\rm p}}\right)^2\right)^{1/2},
\end{equation}
\begin{equation}
 \frac{\delta B_{\rm p}}{B_{\rm p}} = \left(\left(0.10\frac{\delta F_{\rm op}}{F_{\rm op}}\right)^2+
 \left(0.21\frac{\delta D}{D}\right)^2+
 \left(\frac{\delta \nu_{\rm p}}{\nu_{\rm p}}\right)^2\right)^{1/2}.
\end{equation}

\begin{figure}
  \begin{center}
   \includegraphics[scale=1.0]{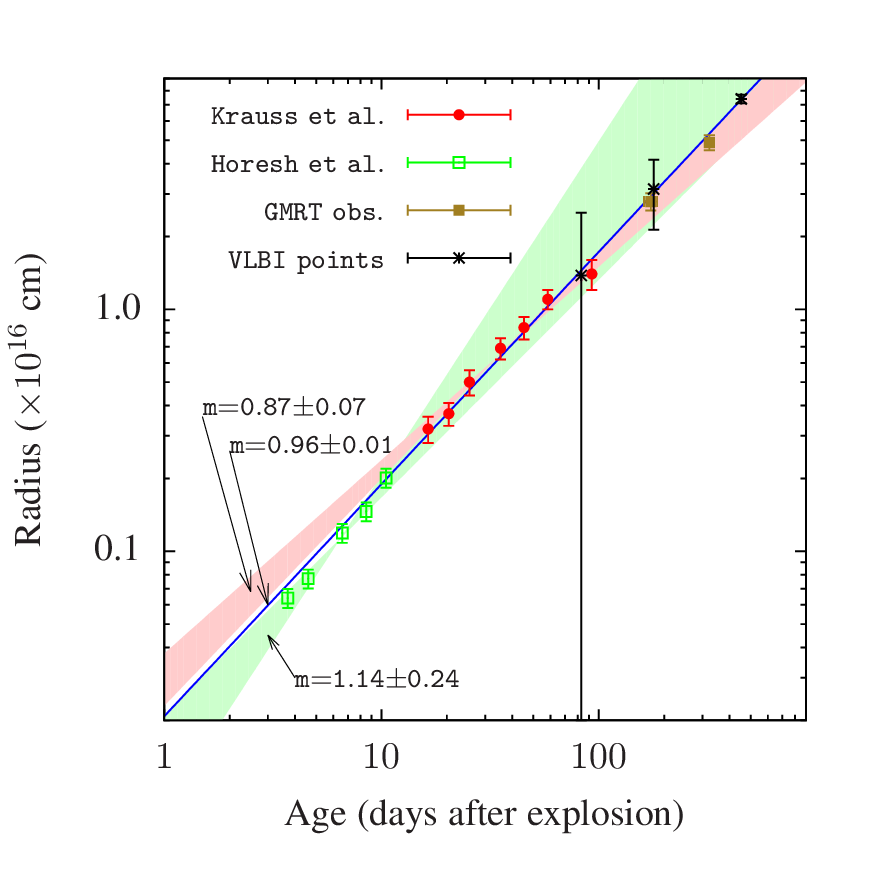}
   \includegraphics[scale=1.0]{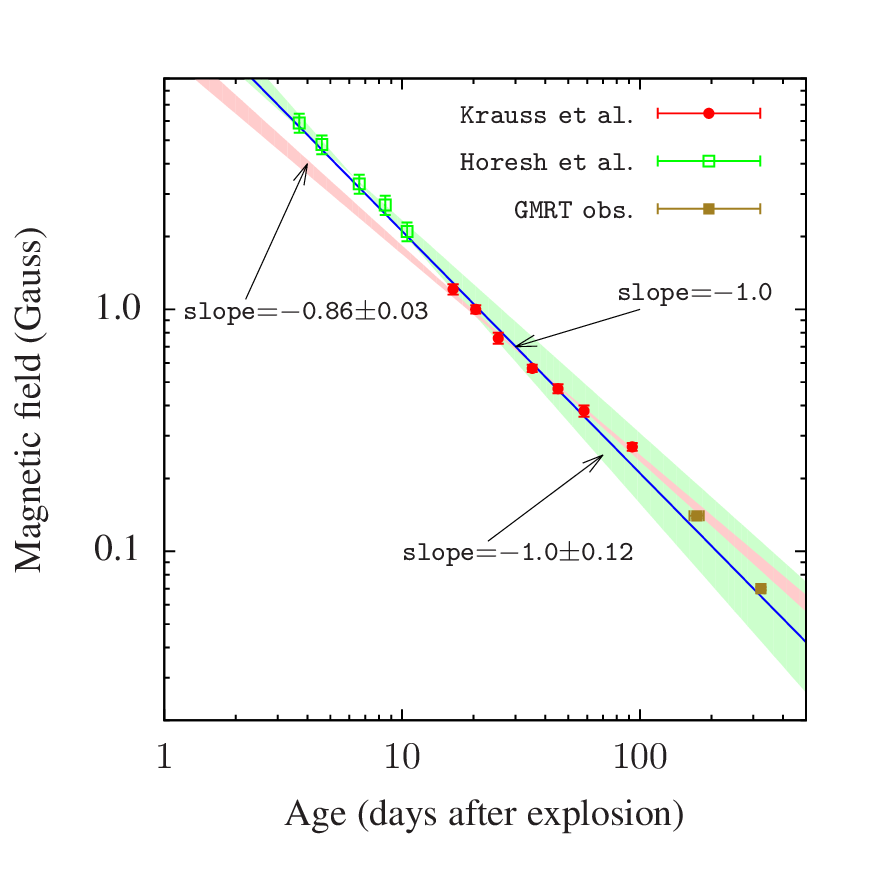}
   \caption{The radius (upper panel) and magnetic field (lower panel) at the two epochs determined using
low frequency data presented along with the values obtained by \citealt{2013MNRAS.436.1258H} using CARMA,
\citealt{2012ApJ...750L..40K} using VLA and by \citealt{2012ApJ...751..125B,2016MNRAS.455..511D} using
Very Large Baseline Interferometry (VLBI) data. The shaded red area corresponds to lines with $m=0.87\pm0.07$
obtained by \citealt{2012ApJ...750L..40K}, while the shades green area with $m = 1.14\pm0.24$ correspond to
the early radio data as obtained by \citealt{2013MNRAS.436.1258H} and the line with $m=0.96\pm0.01$ is based
on value reported in \citealt{2016MNRAS.455..511D}.
}\label{fig:r_and_b}
\end{center}
\end{figure}

\begin{figure}
  \begin{center}
   \includegraphics[scale=1.0]{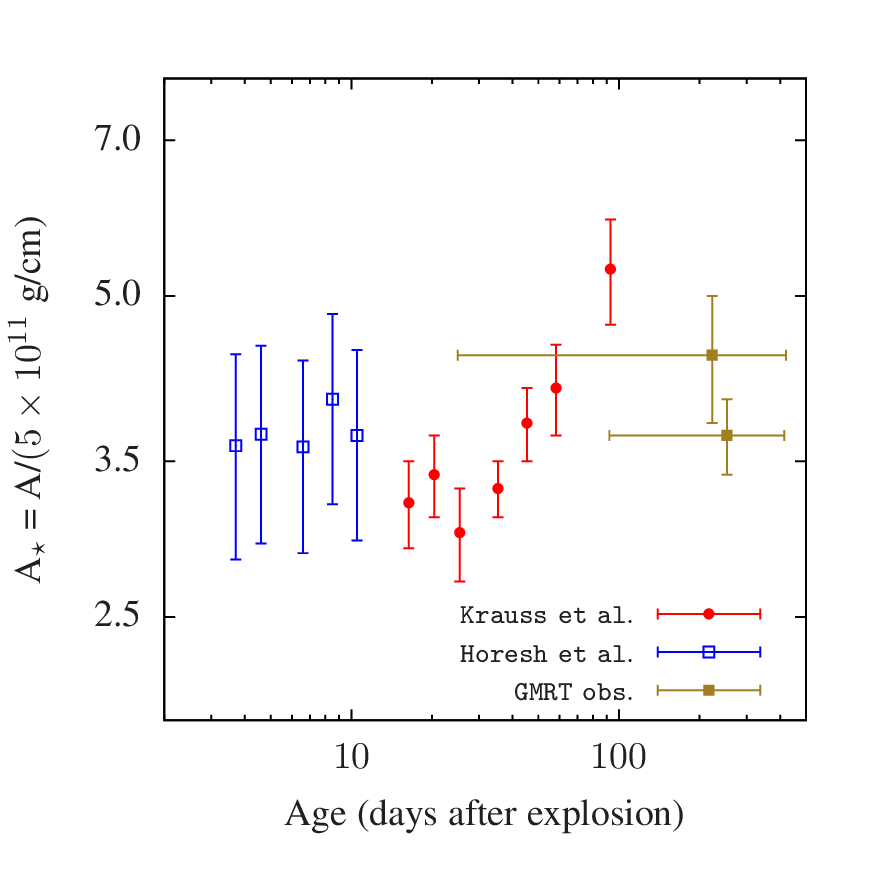}
   \caption{The mass loss parameter $A_{\star}$ plotted for the observation time
range of $3-450\ \rm days$. The parameter shows some variation which is consistent with an
average value of around  $3.5$. All the points are calculated based on
$\epsilon_{\rm B}=0.1$ and are based on fitting a radio spectra except the last two points,
which are based on $610\ \rm MHz$ and $1280\ \rm MHz$ (VLA+GMRT) light curve. The corresponding $A_\star$
is therefore valid for the entire duration of the respective light curve.}\label{fig:A_star}
  \end{center}
\end{figure}

\begin{figure}
  \begin{center}
   \includegraphics[scale=1.0]{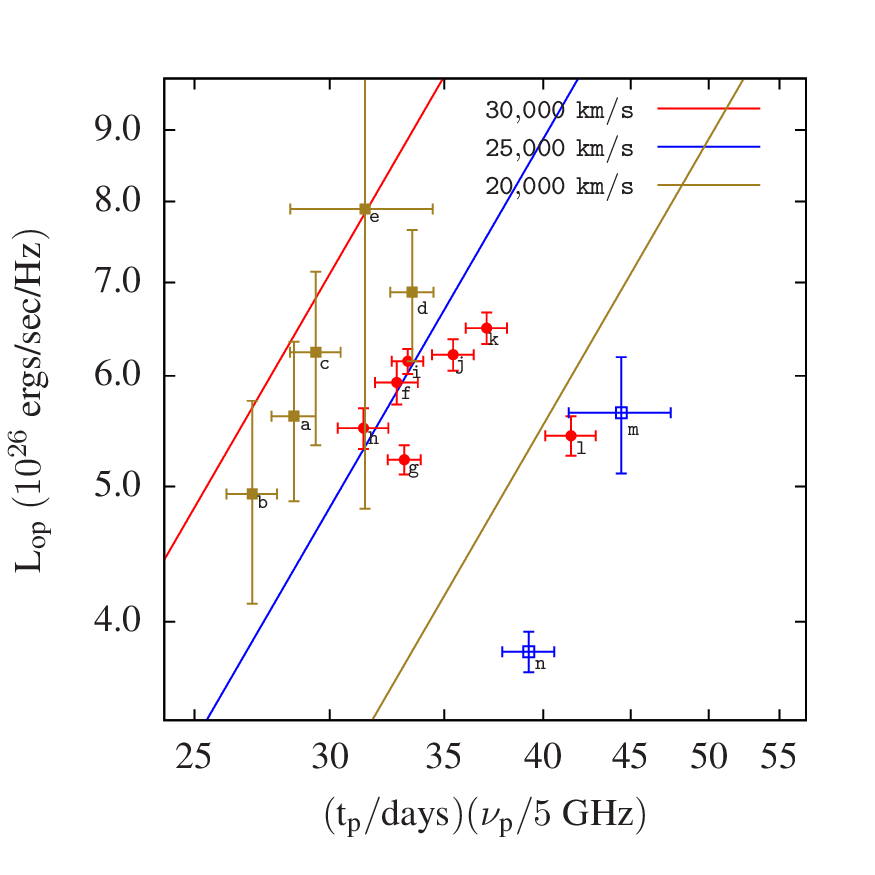}
   \caption{The $L_{\rm op}-\nu_{\rm p}t_{\rm p}$ plot for SN 2011dh showing all the points obtained
   using the fits reported by \citealt{2013MNRAS.436.1258H} (filled squares), \citealt{2012ApJ...750L..40K}
   (filled circles) and the values obtained using low frequency late time data (empty squares). The points
   labeled as [a, b, c, d, e, f, g, h, i, j, k, l, m \& n] correspond to
   [3.7, 4.6, 6.6, 8.5, 10.5, 16.4, 20.4, 25.4, 35.3, 45.3, 58.2, 92.9, 173.6 \& 323.1]
   days respectively. The constant velocity lines have been obtained using Equation~\ref{eqn:chev-diag}
   as in \citealt{1998ApJ...499..810C}. Note the evolution between expansion speeds at early times (VLA) to late times (GMRT).}\label{fig:chev_diag}
  \end{center}
\end{figure}

In Figure~\ref{fig:r_and_b}, we plot the radius and magnetic field values we obtain along with the
values reported in literature
\citep{2012ApJ...750L..40K,2013MNRAS.436.1258H,2012ApJ...751..125B,2016MNRAS.455..511D} for the purpose
of comparison. Our GMRT measurements taken along with VLA and Very long Baseline Interferometry (VLBI)
measurements are consistent with a deceleration index $m= 0.96$ for radio sphere evolution.
The values of dimensionless parameter $A_{\star}$ (using $\epsilon_{\rm B}=0.1$) are plotted in Figure~\ref{fig:A_star}.
All the points are based on fitting a radio spectra except the latest two points, which are based on
$610\ \rm MHz$ and $1280\ \rm MHz$ (VLA+GMRT) light curves, therefore the corresponding $A_\star$ is valid
for the entire duration of the corresponding light curve at a given frequency. There are small variations
in the mass loss rate ($\dot M=6.2\times10^{12}\varv_{\rm w}A_\star\ \rm gm\ s^{-1}$), but the average value
of $A_\star$ is around $\sim 3.7^{+1.3}_{-0.3}$. Using this $A_\star = 3.7$ we obtain a nearly constant
value of mass loss rate of, $\dot M = 7.2 \times 10^{-7} \ \rm M_{\sun}\ yr^{-1}$ for a yellow
supergiant progenitor with an assumed wind speed of $\varv_{\rm w} = 20\ \rm km/s$.
In Figure~\ref{fig:chev_diag}, we plot $F_{\rm op}$ vs. $\nu_{\rm p}t_{\rm p}$ using all the fits from
\citealt{2013MNRAS.436.1258H} and \citealt{2012ApJ...750L..40K} including our measurements based on $610\ \rm MHz$
and $1280\ \rm MHz$. This type of plot was used by \citealt{1998ApJ...499..810C} to distinguish between different
types of supernovae on the basis of their radio emission properties. The peak luminosity $L_{\rm op}$ and mean
forward shock velocity $\varv_{\rm p}$ can be related (for $p=2.8$) using Equation~\ref{eqn:radius} as
\begin{eqnarray}
\varv_{\rm p}&=&3.6\times10^{9}\ f_{\rm eB}^{-0.05}\left(\frac{f}{0.5}\right)^{-0.05}
\left(\frac{L_{\rm op}}{10^{26}\ \rm erg\ s^{-1}\ Hz^{-1}}\right)^{0.47}\nonumber \\
&&\left(\frac{\nu}{5\ \rm GHz}\right)^{-1}
\left(\frac{t_{\rm p}}{10\ \rm days}\right)^{-1}
\ \rm cm\ s^{-1},\label{eqn:chev-diag}
\end{eqnarray}

where $L_{\rm op}=4\pi D^2 F_{\rm op}$ is the peak radio spectral luminosity. We can use it to make
lines for various values of $\varv_{\rm p}$ on a $L_{\rm op}-\nu_{\rm p}t_{\rm p}$ plot which are shown
in Figure~\ref{fig:chev_diag}. We note that object moves on the plot between various constant velocity
lines and the velocity decreases as the object ages. The average shock velocity $2.5\times 10^9\ \rm cm\ s^{-1}$
(based on $f_{\rm eB}=1$) quoted by \citealt{2012ApJ...750L..40K} appears to be on the left corner of this
diagram (refer to Figure~\ref{fig:chev_diag}), while that of the \citealt{2013MNRAS.436.1258H}
overlaps with the range of velocities shown between the black and blue lines. The shock appears to slow down
with time according to this diagram which supports $m<1$.

\begin{table}
\caption{Results of the fit to low frequency GMRT radio observations of SN 2011dh.
The Age is calculated assuming 2011 May 31.8 UT as the explosion
date. The value of $\epsilon_{\rm B}$ is taken to be $0.1$ in calculating $A_{\star}$.}
\centering
 \begin{tabular}{l c c}
 \hline
\rm Frequency & $1280\ \rm MHz$ & $607\ \rm MHz$  \\
\hline
\rm Peak flux density (mJy)& $6.20 \pm 0.16$ & $4.51 \pm 0.23$ \\
\rm Age at peak (days)& $173.6 \pm 11.9$ & $323.1 \pm 11.3$ \\
\rm Radius ($10^{16}\ \rm cm$)  & $2.79\pm 0.23$ & $4.89\pm 0.35$ \\
\rm Magnetic field (Gauss) & $0.135\pm 0.003$ & $0.069\pm 0.001$ \\
\rm $A_{\star}$ & $4.4\pm 0.60$ & $3.72\pm 0.29$ \\
$\chi_\nu^2$ & $4.8$ & $0.8$\\
\hline
\end{tabular}
\label{Table:fit}
\end{table}

\section{Discussions}\label{sec:discussions}
In Figure~\ref{fig:r_and_b} we show the time evolution of the shock wave radius based on GMRT data together
with VLBI and early VLA data. We have drawn a line with slope $m \sim 0.96$ \citep{2016MNRAS.455..511D}, and
the shaded error regions correspond to error in the fits by \citealt{2012ApJ...750L..40K} ($m\sim0.87\pm0.07$) and
\citealt{2013MNRAS.436.1258H} ($m\sim1.14\pm0.24$).
It is evident that the SN shock has been propagating through a circumstellar medium set up by a
wind with constant mass loss parameter, $\dot M/v_{\rm w}$. If we assume that a pre-SN stellar
wind speed, $v_w = 20\ \rm km/s$ (corresponding to the constant wind speed of a yellow supergiant), then the
radio observations imply an average value of mass loss rate of $\dot M = 7.2 \times 10^{-7} \ \rm M_{\sun} yr^{-1}$, which
is a factor of $\sim 4$ smaller than the determination by \citealt{2014ApJ...785...95M} from X-ray analysis.
We note that the deviations of predicted flux densities from the observed radio flux densities is more for
the case of \citealt{2013MNRAS.436.1258H} compared to \citealt{2012ApJ...750L..40K}
(see Figure~\ref{fig:fits} \& \ref{fig:fits325}). \citealt{2013MNRAS.436.1258H} found a value of $m$ which
 is marginally consistent with the physically acceptable
$m$ obtained by \citealt{2012ApJ...750L..40K}, ($m\sim 0. 87 \pm 0.07$) and the former authors' central value of index $m$
was biased on the higher side of unity which is in an improbable dynamical regime given the asymptotic solution due to
\citealt{1982ApJ...258..790C} for a shock wave in medium set up by a wind with constant mass loss rate, has the shock
decelerating as $m = (n - 3)/(n - 2) < 1$, where $n$ is the index for ejecta density profile.
A possible reason may be the change in the properties of the blast wave or
the electron population with time as the \citealt{2013MNRAS.436.1258H} model is based on very early time data
while \citealt{2012ApJ...750L..40K} model is based on slightly late time data.
It is noted that the velocity of blastwave seems to have decreased with time as expected
for an interacting blast wave (shown in Figure~\ref{fig:chev_diag}). The points are located
between constant velocity lines corresponding to  $2.5\times 10^{9}\ \rm cm\ s^{-1}$ and
$3.0\times 10^{9}\ \rm cm\ s^{-1}$ at early time ($<58\ \rm days$), and thereafter ($>90\ \rm days$)
move towards lower values of velocity of around
$2.0 \times 10^{9}\ \rm cm\ s^{-1}$. On the basis of early optical spectra \citealt{2011ApJ...742L..18A} suggested
that the supernovae had a compact progenitor ($R_\star \sim 10^{11}\ \rm cm$). \citealt{2012ApJ...752...78S}
also supported the hypothesis with a compact progenitor star using the initial X-ray and radio data.
On the other hand, a slowed down shock as found in the size estimates of the radio sphere based on synchrotron
spectra from early CARMA, VLA and our late time GMRT data is consistent with a type eIIb classification \citep{2010ApJ...711L..40C}.
Also, the VLBI measurements of \citealt{2016MNRAS.455..511D} confirms that by $453\ \rm days$, the average
expansion velocity was reduced to $1.89 \pm 0.28 \times 10^{9}\ \rm cm \ s^{-1}$ (modulo a slightly different
smaller distance of $\sim 7.8\ \rm Mpc$). Also, type cIIb SNe show late time radio variability as for example in
SN 2001ig \citep{2004MNRAS.349.1093R} or SN 2003bg \citep{2006ApJ...651.1005S} which SN 2011dh did not
display. The fact that the GMRT radio data is also consistent with evolution from the earlier VLA
phase (albeit a shock speed change between $58$ to $92\ \rm days$) points to the extended nature of type
progenitor rather than a compact Wolf-Rayet progenitor.

The residual blue point source found at the location of SN 2011dh
could be due to a combination of the following four possible causes as pointed
out by \citealt{2015MNRAS.454.2580M}, namely: a) a binary companion of the exploded
supernova YSG progenitor, b) light echo from dust around the SN position,
c) SN 2011dh itself (either due to Freeze-out in the Helium envelope or in
combination with radioactivity powered optical emission) and d) due to circumstellar
interaction of the SN. \citealt{2015MNRAS.454.2580M} discard the light echo scenario (case-b)
due to the special arrangement of dust directly behind the SN in the line of sight required
to produce the point-like residual source. They also use their Swift X-ray Telescope
(XRT) and VLA observations to argue against ongoing CSM interaction at the time of late
time HST observations. However, we note that their $5\ \rm ksec$ XRT exposure gives a
$3 \sigma$ upper limit of $1.2 \times 10^{39}\ \rm erg\ s^{-1}$ which is higher than the
predicted X-ray luminosity ($5 \times 10^{37} \ \rm erg\ s^{-1}$ at $500\ \rm days$).
\citep{2012ApJ...752...78S} based on early Swift data and
the measured Chandra luminosity of $5.5 \times 10^{37}\ \rm erg\ s^{-1}$ which is much
lower even at $500\ \rm days$ \citep{2014ApJ...785...95M} i.e. half the age of the SN considered
by \citealt{2015MNRAS.454.2580M}. Therefore the X-ray observations do not rule out any ongoing
strong CSM interaction continuing till $1164\ \rm days$ as claimed by \citealt{2015MNRAS.454.2580M}.
They further claim that their VLA observation on October 18, 2014 UT (day $1236$) which showed a
radio flux density of $459 \pm 93 \ \rm \mu Jy$ in the C-band ($6.1\ \rm GHz$)
and $109 \pm 17 \ \rm \mu Jy$ in the K-band ($22\ \rm GHz$) are well within the expectations
based on earlier VLA measurements, assuming that the CSM has an extended wind-like structure.
As already argued here, the \citealt{2013MNRAS.436.1258H} predictions somewhat overestimate the
optically thin part of the emission in the L-band even during our GMRT observation epochs.
In fact, with our GMRT L-band ($1.387\ \rm GHz$) measurements on day $421$, we would expect about
$230 \ \rm \mu Jy$ at $6.1\ \rm GHz$ and on day $1229$ using
Equation~\ref{eqn:Horesh1} \& \ref{eqn:Horesh2}
which is about half of what has been reported. This, together with the evidence of early
slowing down of the shock both in our GMRT data as well as in the long term average of VLBI
measurements show that the there is fairly strong ongoing CSM interaction which may have even
been enhanced by day $1229$ and must be contributing substantially in other bands
(optical and UV included) within the as yet point source. There is also evidence that the
nebular line profiles of SN 2011dh observed between $201$ and $678\ \rm days$ show a roughly
spherical explosion with a few aspherical clumps and the SN is showing signs of strong CSM
interaction \citep{2013MNRAS.436.3614S}. However, the CSM interaction may decrease with time
if the circumstellar wind is roughly spherically symmetric and with time may even be resolved
if it continues to remain strong.

A comparison with the prototypical type IIb SN 1993J which had a
considerable fraction of its hydrogen envelope stripped off prior to explosion
shows that SN 2011dh had less radio luminosity at its peak and it evolved
much more rapidly compared to SN 1993J (see \citealt{2016MNRAS.455..511D} for a comparison).
This fast rise to peak of the radio light curve as well as the low peak luminosity imply
that the CSM surrounding SN 2011dh was less dense than that for SN 1993J.
There is thus variation of the properties of the environment and past history of the
progenitors of the same spectroscopic type of SN.
\section{Conclusions}
We find that the radio emission models based on early time data ($<16\rm \ days$) of SN 2011dh
are not very suitable for predicting the late time ($>100 \ \rm days$) radio flux densities. This
may be because the parameters of the shock-CSM interaction may have varied over time as exemplified
by e.g. the variation of shock wave velocity with time. Such a
hypothesis will be the scope of future investigations in similar types of SNe. We note that the SN 2011dh
has many unsettled questions regarding binarity, mass and mass loss of the progenitor which if probed will
go a long way in helping us understand supernovae and their progenitors in general.
\section{Acknowledgments}
We wish to acknowledge the support of TIFR 12th Five Year Plan (Project No: 12P-0261). NY wishes to
thank Asaf Horesh for his valuable comments on the manuscript. NY wishes to acknowledge the support
of CSIR-SPM fellowship (SPM-07/858(0057)/2009-EMR-I). AR thanks the director and staff of ITC, Harvard
university for their hospitality during his leave of absence from TIFR. We thank the staff of the GMRT who have made
these observations possible. GMRT is run by the National Centre for Radio Astrophysics of the Tata
Institute of Fundamental Research.

\bibliographystyle{mnras}
\bibliography{references}

\end{document}